\documentclass{iau}
\usepackage{graphicx}

\title[JD .~~Bp stars in Orion OB1 association] 
{Bp stars in Orion OB1 association}

\author[Iosif I. Romanyuk \& Ilya A. Yakunin]   
{Iosif I. Romanyuk$^1$
 \and Ilya A. Yakunin $^1$}

\affiliation{$^1$Special Astrophysical Observatory of Russian Academy of Sciences,\newline
369167, Nizhny Arkhyz, Russia \\ email: {\tt roman@sao.ru}}

\pubyear{2013}
\volume{302}  
\pagerange{   -- }
\jname{Magnetic fields throughout stellar evolution}
\editors{A.C. Editor, B.D. Editor \& C.E. Editor, eds.}
\begin{document}

\maketitle

\begin{abstract}
A total of 85 CP stars of various types are identified among 814 members
of the Orion OB1 association. We selected 59 Bp stars, which account
for 13.4\% of the total number of B type stars in the association.
The fraction of peculiar B type stars in the association is found to be
twice higher than that of peculiar A type stars.

Magnetic field are found in 22 stars, 17 of them are objects with anomalous
helium lines.
No significant differences are found between the field strengths in the
Bp type stars of the association and  Bp type field stars.
We identified 17 binaries, which make up 20\% of the total number
of peculiar stars studied  which is the standard ratio for CP stars.
\end{abstract}

\firstsection 
\section{Introduction}

The Orion constellation hosts one of the most popular groups of early-type
stars in the solar neighborhood - the Orion OB1 association.
\cite[Blaau (1964)]{Blaau64}
identified four regions inside the association -
subgroups: a (corresponds to the northen part), b (the Orion's Belt),
c (the region located south of Orion's Belt), and  d (the very compact region
located in the central part of the association) that slightly differ  in age and stellar
composition.

Most of the objects in the Orion OB1 association are normal hot main sequence
stars; however, the association also includes pre-MS objects, like HAEBE stars,
T Tau-type stars, and various anomalous (peculiar) stars.  Chemically peculiar
(CP) stars differ from normal stars by their anomalous chemical composition  which
shows up in enhanced or weakened intensity of lines of certain elements
in the stellar spectrum.

\cite[Renson and Manfroid (2009)]{renson09} published the most detailed
catalog of CP stars, which includes more than 8200 objects.
Over the past quarter-century many new observations of CP stars have been
performed. Our aim is to thoroughly analyze massive chemically peculiar and
magnetic stars in the Ori OB1 association using all available data. For a review
and analysis of the main studies on this subject, see
\cite[Romanyuk \& Yakunin (2012)]{RomYak12} and
\cite[Romanyuk et al., (2013)]{Rom_etal13}.

\section{CP stars in the association}

Groups of hot stars in the Ori OB1 association have repeatedly attracted the
researcher's attention. Here we consider only the issues related to chemically
peculiar stars and magnetic field of these objects.

\cite[Borra \& Landstreet (1979)]{BoLa79}
discovered very strong magnetic fields in a group
of B-type stars with enhanced helium lines in young clusters in Orion.
\cite[Klochkova (1985)]{Klo85}
performed spectroscopic observations of 24 CP stars to determine
the distance moduli and ages of subroups.
\cite[Brown et al. (1994)]{Brown94}
reported the results of photometric observations in the Walraven system
for 814 stars, for all identified of suspected association members.
They determined the effective temperatures, surface gravities, luminosities
and masses for all 814 stars. They also detemined the distance moduli and
showed that the near and far edges of clouds in the Orion OB1 association are
located at the distances of about 320 and 500 pc, respectively.
We decided to identify  chemically peculiar stars among the 814 objects of
this list. We consider a star to be peculiar if it is appeares in the
catalog by
\cite[Renson and Manfroid (2009)]{renson09}.

We selected 85 CP stars in the direction of the Ori OB1 association.
We list these objects in paper by
\cite[Romanyuk et al., (2013)]{Rom_etal13}. Most of them
(59 objects) are Bp stars, however, we also found 23 Am and 3 Ap stars.
We performed magnetic field observations using the 6m Russian telescope.

The age of subgroups, number of normal and CP stars in each subgroup
and fraction of CP stars are presented in Table \ref{tab1}.

\begin{table}
  \begin{center}
  \caption{Age of subgroups and number of normal and CP stars }
  \label{tab1}
{\scriptsize
  \begin{tabular}{|l|c|c|c|c|}
\hline
subgroup & age, log t &  all stars & CP stars &   fraction, \%  \\
\hline
Ori OB1a &    7.05   &   311    &     24   &      7.7   \\
Ori OB1b &    6.23   &   139    &     21   &     15.1    \\
Ori OB1c &    6.66   &   350    &     37   &     10.6     \\
Ori OB1d &    6.0    &    14    &      3   &     21.4     \\
\hline
  \end{tabular}
   }
 \end{center}

\end{table}

We determine distances for most of the stars closer than 250 pc
from their Hipparcos parallaxes and we estimate the distances
to more distant objects from their temperatures and luminosities.
Proper motions have been measured for all CP stars.

The sample of peculiar stars is offset relative to the entire sample both
in terms of temperature and luminosity. The fraction of hot stars
is greater among peculiar stars. The effective-temperature distribution
for the entire sample and CP stars peak at $\log T_e$ =3.95 and
$\log T_e$ =4.15 (for details  see
\cite[Romanyuk et al., (2013)]{Rom_etal13}.
The distribution of CP stars in different subroups is presented
in Table \ref{tab2}.

\begin{table}
  \begin{center}
  \caption{Number of CP stars in different subroups }
  \label{tab2}
{\scriptsize
  \begin{tabular}{|l|c|c|c|c|c|}
\hline
Peculiarity type & total &  Ori OB1a &  Ori OB1b &  Ori OB1c &  Ori OB1d  \\
\hline
Am        &   23     &    6    &     4   &   13   &   0   \\
He-strong &    7     &    1    &     3   &    1   &   2   \\
He-weak   &   27     &    7    &     8   &   12   &   0   \\
Si, Si+   &   19     &    6    &     4   &    8   &   0   \\
Other     &    9     &    3    &     1   &    4   &   1  \\
\hline
  \end{tabular}
   }
 \end{center}

\end{table}

The catalog of CP stars by
\cite[Renson and Manfroid (2009)]{renson09} includes 23 Am stars
in the directions of the association.
This is surprising given that low-mass Am stars should not have
yet evolved enough to settle onto the main sequence.
We therefore decided to verify whether the Am stars in question are
foreground objects and not members of the association.
Parallaxes are available for 14 of the 23 Am stars and they support
conclusively the above hypothesis - these objects are located
closer than 300 pc. The distances of remaining nine stars can
be determined only from analysis of their tempearures and luminosities.
The result of analysis is the same: the distances to Am stars are closer
than 300 pc.

\begin{figure}[b]
\begin{center}
 \includegraphics[width=3.4in]{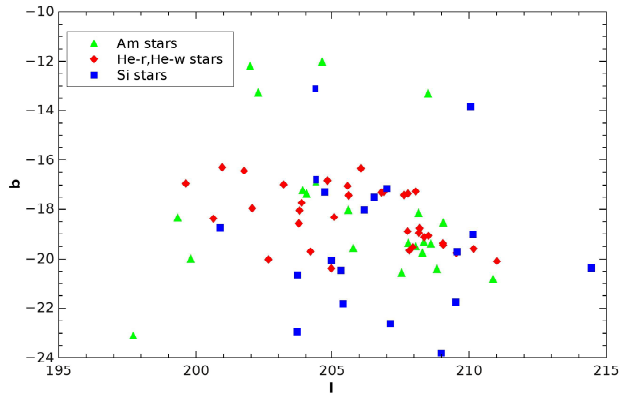}
 \caption{Spatial distribution of CP stars in the Ori OB1 assocation.}
   \label{fig1}
\end{center}
\end{figure}

{\underline{\it Occurence frequency of CP stars }}.

The fraction of CP stars can be seen to be smallest (7.7\%) in the
oldest subgroup (a) of the association. It is twice higher (15.1\%) in the
substantially younger subgroup (b). The fraction of peculiar stars
is even higher in the youngest subgroup (d), however, it contains
too few (only 14) objects to allow any statistical conclusions.
The lists of
\cite[Brown et al (1994)]{Brown94} contains 661 stars in the two older
groups (a) and (c): 81.2\% of the total number of stars in the
association. We adopt this fraction as standard.
Bp stars are much less concentrated in old subroups. Thus the fraction
of He-weak stars in the old subgroups (a) and (c) is equal to 70\%;
that of Si-stars, to 52.6\%, and that of He-strong stars, to 28.6\%.
The discrepancies are quite substantial and significant.
The predominance of stars with enhanced helium lines is especially
conspicuous in the young subgroups (b) and (d). It is interesting that the
fraction of He-strong stars in subgroups (b) and (d), whose ages do not
exceed 2 Myr, is three times higher than that of He-weak stars in the
same subgroups.

{\underline{\it Spatial distribution of CP stars }}.

Spatial distribution of CP stars in the Orion OB1 association is demonstrated
on Fig.\,\ref{fig1}.

The group of Am stars is located separately from the other objects and closer to us.
It can be concluded that Am stars are foreground objects  which are merely seen projected
against the association.
Stars with anomalous helium lines belongs to cluster and mostly concentrate
in subgroups (b) and (c) - 20 objects, whereas subgroup (a) contains only
8 objects. Silicon stars concentrate mostly in subgroups (a) and (c) -
14 out of 18 objects.

{\underline{\it Binary CP stars}}.

We identified 17 physical binaries among 85 CP stars of the association, which
makes up a typical fraction of 20\%.
These stars are distributed by peculiarity type as follows: He-strong - 5 (out
of 7). He-weak - 4 (out of 27), SiSi+ - 2 (out of 10), Am - 1 (out of 23), and
other types -  5 (out of 9).

The largest fraction of binaries is found among He-strong stars. The two stars
with no companion found are HD\,36982 and HD\,37776.
Weak Am stars are very poorly studied, and abnormally low number of binaries
may be result of absence of radial velocity measurements.
He-weak and  SiSi+ stars show the fraction of binaries lower than normal.

\section{Magnetic fields}

We found 22 magnetic stars in the association of which 21 are  Bp stars.
Table \ref{tab3} lists the stars with reliably detected magnetic field.

\begin{table}
  \begin{center}
  \caption{Magnetic stars in the Ori OB1 association }
  \label{tab3}
{\scriptsize
  \begin{tabular}{|l|l|c|c|l|l|c|}
\hline
Star & Sp  Pec       & $B_e$ extrema &  & Star  & Sp  Pec & $B_e$ extrema  \\
\hline
HD 35008 &  B8  Si     & -340        &   & HD 36668 &  B7  He-wk, Si & -2200/+2000\\
HD 35298 &  B6  He-wk  & -3000/+3000 &   & HD 36916 &  B8  He-wk, Si & -1100/0   \\
HD 35456 &  B7  He-wk  &  -400/+1080 &   & HD 36955 &  A2  CrEu      & -1300/-410 \\
HD 35502 &  B6  SrCrSi & -2250/-180  &   & HD 37017 &  B2  He-strong & -2300/-300 \\
HD 35730 &  B7  He-wk  &  -450/+250  &   & HD 37058 &  B2  He-wk, Sr & -1200/+1200\\
HD 36313 &  B8  He-wk  & -1500/-1100 &   & HD 37140 &  B8  SiSr      & -1050/+400\\
HD 36429 &  B6  He-wk  &  -840/+160  &   & HD 37479 &  B2  He-strong & -1600/+3500\\
HD 36485 &  B2  He-strong&-3700/+3000&   & HD 37642 &  B9  He-wk, Si & -3000/+3000\\
HD 36526 &  B8  He-wk,Si&-3500/+3400 &   & HD 37687 &  B7  He-wk     & -600/+500\\
HD 36540 &  B7  He-wk  &  -900/+1030 &   & HD 37776 &  B2  He-strong & -2000/+2000\\
HD 36629 &  B3  He-wk  &  -1300/+1100&   & HD 290665&  B9  SrCrEu    & -1600/+5000\\
\hline
  \end{tabular}
   }
 \end{center}

\end{table}

Eight of the magnetic stars are binary (36.4\%). The overwhelming
majority (17 out of 22, or 77\%) are objects with anomalous
helium lines. The fraction of magnetic stars in the inner subgroup (b)
is twice higher than in the outer subroups (a) and (c).

We see no significant differences between magnetic stars of the Ori OB1
association and Bp stars in general in terms of magnetic field strength.
However, despite poor statistics, He-strong stars can be seen to posses,
on the whole, a factor of 1.5 -2 stronger fields than He-wk stars.

\section{Conclusion}

We thus identified 85 CP stars in the direction toward the young Ori OB1
association. Our CP stars are distributed by peculiarity
types as follows: 23 Am stars, 3 Ap stars and 59 Bp stars.
The fraction of peculiar B-type stars in the association is twice higher
than the corresponding fraction of peculiar A-type stars.
The association includes 22 magnetic stars; 21 of them  are Bp stars, and
only one is an Ap star. We suggest that when the stars were born in the Orion OB1
association the magnetic fields formed mostly in the objects that
later developed helium rather than silicon anomalies.

\begin{discussion}

\discuss{Khalak}{ Have you found any correlation between the spatial distribution
stars with the magnetic field and the structure of galactic magnetic field in the area
nearby Orion OB1 association? }

\discuss{Romanyuk}{ The number of known magnetic stars are too small
 to look for any correlation. The study of magnetic stars in Orion is the first step
in this direction.}

\discuss{Wade}{Can you measure radial velocities of stars from your spectra ?}

\discuss{Romanyuk}{Yes, of course. We have spectra with the resolution R = 15000
and R=40000 and SNR = 200-300. So we can look for binaries among stars in
the Ori OB1 association.}

\end{discussion}


\begin{thebibliography}{}

\bibitem[Blaau (1964)]{Blaau64}
{Blaau A.}, 1964, \textit{ARAA}, 2, 236.

\bibitem[Romanyuk \& Yakunin (2012)]{RomYak12}
{Romanyuk I.I., Yakunin I.A.}, 2012, \textit{Astrophys Bull.}, 67, 177.

\bibitem[Romanyuk et al., (2013)]{Rom_etal13}
{Romanyuk I.I. et al.}, 2013, \textit{Astrophys Bull.}, 68, 300.


\bibitem[Borra \& Landstreet (1979)]{BoLa79}
{Borra E.F., Landstreet J.D.}, 1979, \textit{A\&A}, 228, 809.

\bibitem[Klochkova (1985)]{Klo85}
{Klochkova V.G.}, 1985, \textit{PAZ}, 11,209.

\bibitem[Brown et al (1994)]{Brown94}
{Brown A.G.A. et al.}, 1994, \textit{A\&A}, 289, 101.


\bibitem[Renson and Manfroid (2009)]{renson09}
{Renson P., Manfroid J.}, 2009, \textit{A\&A}, 498, 961.


\end{thebibliography}
\end{document}